\documentclass[prd,superscriptaddress,unsortedaddress,
twocolumn,showpacs,preprintnumbers,amsmath,amssymb]{revtex4}

\usepackage{graphicx}
\usepackage{amsmath,amssymb,times}
\usepackage{braket}

\def\fun#1#2{\lower3.6pt\vbox{\baselineskip0pt\lineskip.9pt
\ialign{$\mathsurround=0pt#1\hfil##\hfil$\crcr#2\crcr\sim\crcr}}}

\newcommand{\beq}{\begin{equation}}
\newcommand{\eeq}{\end{equation}}
\newcommand{\bea}{\begin{eqnarray}}
\newcommand{\eea}{\end{eqnarray}}

\DeclareSymbolFont{boldletters}{OML}{cmm} {b}{it}
\DeclareSymbolFontAlphabet{\mathbit}{boldletters}
\DeclareMathSymbol{\alpha}{\mathalpha}{letters}{"0B}
\DeclareMathSymbol{\beta}{\mathalpha}{letters}{"0C}
\DeclareMathSymbol{\gamma}{\mathalpha}{letters}{"0D}
\DeclareMathSymbol{\delta}{\mathalpha}{letters}{"0E}
\DeclareMathSymbol{\epsilon}{\mathalpha}{letters}{"0F}
\DeclareMathSymbol{\zeta}{\mathalpha}{letters}{"10}
\DeclareMathSymbol{\eta}{\mathalpha}{letters}{"11}
\DeclareMathSymbol{\theta}{\mathalpha}{letters}{"12}
\DeclareMathSymbol{\iota}{\mathalpha}{letters}{"13}
\DeclareMathSymbol{\kappa}{\mathalpha}{letters}{"14}
\DeclareMathSymbol{\lambda}{\mathalpha}{letters}{"15}
\DeclareMathSymbol{\mu}{\mathalpha}{letters}{"16}
\DeclareMathSymbol{\nu}{\mathalpha}{letters}{"17}
\DeclareMathSymbol{\xi}{\mathalpha}{letters}{"18}
\DeclareMathSymbol{\pi}{\mathalpha}{letters}{"19}
\DeclareMathSymbol{\rho}{\mathalpha}{letters}{"1A}
\DeclareMathSymbol{\sigma}{\mathalpha}{letters}{"1B}
\DeclareMathSymbol{\tau}{\mathalpha}{letters}{"1C}
\DeclareMathSymbol{\upsilon}{\mathalpha}{letters}{"1D}
\DeclareMathSymbol{\phi}{\mathalpha}{letters}{"1E}
\DeclareMathSymbol{\chi}{\mathalpha}{letters}{"1F}
\DeclareMathSymbol{\psi}{\mathalpha}{letters}{"20}
\DeclareMathSymbol{\omega}{\mathalpha}{letters}{"21}
\DeclareMathSymbol{\varepsilon}{\mathalpha}{letters}{"22}
\DeclareMathSymbol{\vartheta}{\mathalpha}{letters}{"23}
\DeclareMathSymbol{\varpi}{\mathalpha}{letters}{"24}
\DeclareMathSymbol{\varrho}{\mathalpha}{letters}{"25}
\DeclareMathSymbol{\varsigma}{\mathalpha}{letters}{"26}
\DeclareMathSymbol{\varphi}{\mathalpha}{letters}{"27}
\DeclareMathSymbol{\Gamma}{\mathalpha}{letters}{"00}
\DeclareMathSymbol{\Delta}{\mathalpha}{letters}{"01}
\DeclareMathSymbol{\Theta}{\mathalpha}{letters}{"02}
\DeclareMathSymbol{\Lambda}{\mathalpha}{letters}{"03}
\DeclareMathSymbol{\Xi}{\mathalpha}{letters}{"04}
\DeclareMathSymbol{\Pi}{\mathalpha}{letters}{"05}
\DeclareMathSymbol{\Sigma}{\mathalpha}{letters}{"06}
\DeclareMathSymbol{\Upsilon}{\mathalpha}{letters}{"07}
\DeclareMathSymbol{\Phi}{\mathalpha}{letters}{"08}
\DeclareMathSymbol{\Psi}{\mathalpha}{letters}{"09}
\DeclareMathSymbol{\Omega}{\mathalpha}{letters}{"0A}

\begin{document}

\title{
Properties of 2+1-flavor QCD in the imaginary chemical potential region: 
model approach
}

\author{Junpei Sugano}
\email[]{sugano@phys.kyushu-u.ac.jp}
\affiliation{Department of Physics, Graduate School of Sciences, Kyushu University,
             Fukuoka 819-0395, Japan}             

\author{Hiroaki Kouno}
\email[]{kounoh@cc.saga-u.ac.jp}
\affiliation{Department of Physics, Saga University,
             Saga 840-8502, Japan}  

\author{Masanobu Yahiro}
\email[]{yahiro@phys.kyushu-u.ac.jp}
\affiliation{Department of Physics, Graduate School of Sciences, Kyushu University,
             Fukuoka 819-0395, Japan}

\date{\today}

 \begin{abstract}
  We study properties of 2+1-flavor QCD
  in the imaginary chemical potential region by using two 
  approaches. One is a theoretical approach based on
  QCD partition function,
  and the other is a qualitative one based on
  the Polyakov-loop extended Nambu--Jona-Lasinio (PNJL) model. 
  In the theoretical approach, we clarify conditions imposed on
  the imaginary chemical potentials $\mu_{f}=i\theta_{f}T$
  to realize the Roberge-Weiss (RW) periodicity.
  Here, $T$ is temperature, the index $f$ denotes the flavor,
  and $\theta_{f}$ are dimensionless chemical potentials.
  We also show that the RW periodicity is broken
  if anyone of $\theta_{f}$ is fixed to a constant value.
  In order to visualize the condition, we use the PNJL model 
  as a model possessing the RW periodicity, and  
  draw the phase diagram as a function of $\theta_{u}=\theta_{d}\equiv \theta_{l}$
  for two conditions of $\theta_{s}=\theta_{l}$ and $\theta_{s}=0$. 
  We also consider two cases,
  $(\mu_{u},\mu_{d},\mu_{s}) =(i\theta_{u}T,iC_{1}T,0)$ and
  $(\mu_{u},\mu_{d},\mu_{s})=(iC_{2}T,iC_{2}T,i\theta_{s}T)$;
  here $C_{1}$ and $C_{2}$ are dimensionless constants, whereas 
  $\theta_{u}$ and $\theta_{s}$ are treated as variables.
  For some choice of $C_{1}$ ($C_{2}$),
  the number density of up (strange) quark becomes smooth
  in the entire region of $\theta_{u}$ ($\theta_{s}$)
  even in high $T$ region.
  This property may be important for lattice QCD simulations 
  in the imaginary chemical potential region, 
  since it makes the analytic continuation more feasible.
 \end{abstract}

 \pacs{11.30.Rd, 12.40.-y}
 
\maketitle

\section{INTRODUCTION}
One of the most important issues in hadron physics is to clarify
properties of quark matter in finite
temperature and/or quark chemical potential.
The knowledge of thermodynamics
on quark matter is essential to understand
structure of the QCD phase diagram.
As the review of the QCD phase diagram,
see Refs.
\cite{Stephanov, Munzinger-Wambach, Fukushima-Hatsuda, Fukushima-Sasaki}
and references therein.

Lattice QCD (LQCD) simulations may be the most
promising and powerful theoretical tool of investigating
the QCD phase diagram.
As for isospin-symmetric 2-flavor QCD, the fermion matrix
is written as
\begin{eqnarray}
 \mathcal{M}(\mu_{l}) = \gamma_{\mu}D_{\mu}+m_{l}-\gamma_{4}\mu_{l},
  \label{2-flavor_determinant}
\end{eqnarray}
and satisfies $\gamma_{5}$-hermiticity,
 $(\mathcal{M}(\mu_{l}))^{\dag}=\gamma_{5}\mathcal{M}(-\mu_{l})\gamma_{5}$.
Here, $\mu_{l}$ and $m_{l}$ are the light-quark chemical potential and
its mass, respectively.
LQCD simulations are feasible for $\mu_{l}=0$
since $\textrm{det}\mathcal{M}(0)$
is real and positive definite.
However,
the fermion determinant becomes complex
in finite $\mu_{l}$ because
 $\left(\textrm{det}\mathcal{M}(\mu_{l})\right)^{\ast}=\textrm{det}
  \mathcal{M}(-\mu_{l})\ne \textrm{det}\mathcal{M}(\mu_{l})$
  from the $\gamma_{5}$-hermiticity.
This is the well-known sign problem and
makes the importance-sampling method unfeasible.

One of ideas to circumvent the sign problem
is the imaginary chemical potential $\mu_{l}=i\theta_{l}T$,
where $T$ is temperature and $\theta_{l}$ is a
dimensionless chemical potential.
Indeed, the relation
\begin{eqnarray}
(\mathcal{M}(i\theta_{l}T))^{\dag}
=\gamma_{5} \mathcal{M}(i\theta_{l}T)\gamma_{5}
\end{eqnarray}
can be obtained
and hence there is no sign problem,
and positivity of the fermion determinant is also ensured.
From the imaginary $\mu_{l}$ region,
one can extract information of the real $\mu_{l}$ region
by the analytic continuation.
In fact, this approach was successful for the 2-flavor QCD
\cite{Forcrand-Philipsen1, DElia-Lombard,
Wu-Luo-Chen,DElia-Sanfilippo, Forcrand-Philipsen2,
Nagata-Nakamura, Cea_two_flavor, Takahashi1, Bonati_chiral_transition, Takahashi2}.

\begin{figure}[]
  \begin{center}
\includegraphics[width=0.4\textwidth]{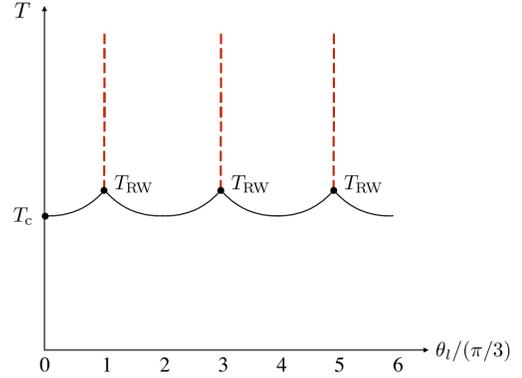}
\end{center}
\caption{
  Sketch of the QCD phase diagram in $\theta_{l}$-$T$ plane.
  The solid line is the crossover deconfinement transition line and
  the vertical dashed line is the first-order RW phase transition line.
  The deconfinement transition temperature is represented by $T_{\rm c}$.
  The label $T_{\rm RW}$ means the RW phase transition temperature.
}
\label{phase_diagram_sketch}
 \end{figure}

In the imaginary $\mu_{l}$ region,
the QCD thermodynamic potential has
the Roberge-Weiss (RW) periodicity~\cite{bib_Roberge-Weiss},
which can be regarded as a
remnant of $\mathbb{Z}_{3}$ symmetry in the pure gauge limit.
Also in Ref.~\cite{bib_Roberge-Weiss},
it was shown that
the first-order RW phase transition occurs at $\theta_{l}=(2k+1)\pi/3$
above some temperature $T_{\rm RW}$,
where $k$ is any integer; see Fig.~\ref{phase_diagram_sketch}.
Due to the RW phase transition,
information of the real $\mu_{l}$ region is limited up to $\mu_{l}/T\sim 1$,
particularly at $T>T_{\rm RW}$.

As an alternative method of LQCD simulations,
one can consider effective models.
Among effective models,
the Polyakov-loop extended Nambu--Jona-Lasinio (PNJL) model
is one of the most useful models
and yields good description of phenomena on quark matter,
such as chiral and deconfinement transitions
\cite{Meisinger, Dumitru, Fukushima1, Ghos, Megias, Ratti1, Rossner,
Kashiwa_PNJL, Sakai_PRD77_051901,
Sakai_PRD78_036001, Sakai_PRD78_076007, Sakai_PRD79_096001, Sakai_JPhys}.
It was proven in Refs.~\cite{Sakai_PRD77_051901,
Sakai_PRD78_036001, Sakai_PRD78_076007, Sakai_PRD79_096001}
that the thermodynamic potential of the PNJL model
possesses the RW periodicity for the 2-flavor case,
and the PNJL  model reasonably 
reproduces LQCD data on 
the imaginary $\mu_l$ region~\cite{Sakai_PRD79_096001, Sakai_JPhys}.  

In the case of 2+1-flavor QCD,
the strange-quark chemical potential $\mu_{s}$
is introduced as an additional external parameter,
and the fermion determinant
consists of the product
$\textrm{det}\mathcal{M}(\mu_{l})\cdot\textrm{det}\mathcal{M}(\mu_{s})$.
When both $\mu_{l}$ and $\mu_{s}$ are pure imaginary,
that is, when $\mu_{l}=i\theta_{l}T$ and $\mu_{s}=i\theta_{s}T$,
the fermion determinant becomes real and
positivity of its determinant is guaranteed just as in the 2-flavor case.
Here, $\theta_{s}$ is a dimensionless chemical potential
for strange quark.
It is thus suitable to consider the imaginary chemical potential region
even in the 2+1-flavor case,
and some works were carried out
\cite{Cea, Bonati, Bonati2, Bonati3,DElia-Gagliardi-Sanfilippo}.
In Ref~\cite{Bonati},
the one-loop effective potential for
the untraced Polyakov loop in the high $T$ limit
was calculated as a function of $\theta_{l}$ for two conditions,
(I) $\theta_{s}=\theta_{l}$ and (II) $\theta_{s}=0$,
and they showed that the RW periodicity exists only in condition (I).
In addition to this result,
the calculation in non-perturbative region
is also necessary to acquire better understanding of
the RW phase transition.

Also in Ref.~\cite{Bonati}, it was pointed out that
the $\theta_{l}$ region
available for analytic continuation becomes broader in condition (II) than (I).
This fact indicates that the analytic region can be expanded
by breaking the RW periodicity deliberately.
It is, therefore, interesting to consider how largely the
analytic region is expanded by breaking the RW periodicity.

In this paper, we study properties of the 2+1-flavor QCD
in the imaginary chemical potential region by using two 
approaches. One is a theoretical approach based
on the QCD partition function,
and the other is a qualitative one based on  the PNJL model.
In the theoretical approach, we first prove
that the thermodynamic potential of 
non-degenerate three-flavor QCD has the RW periodicity in general,
but the periodicity is lost when
anyone of the chemical potentials is fixed to a constant value.
Next, as for the 2+1-flavor case,
we prove that the thermodynamic potential
of the PNJL model has the same properties of QCD
on the RW periodicity.
For this reason,
the PNJL model is used for qualitative analysis.
We calculate some thermodynamic quantities
and draw the phase diagram by 
using the PNJL model
under conditions (I) and (II) in order to visualize 
roles of the conditions.
Finally, we evaluate up- and strange-quark number densities
for some choices of $\theta_{l}$ and $\theta_{s}$.
We numerically confirm that discontinuity of number densities
due to the first-order phase transition disappears in high $T$ region,
and the number densities become smooth. 
This property may be important for LQCD simulations 
in the imaginary chemical potential region, 
since it makes the analytic continuation more feasible
even in high $T$ region.

This paper is organized as follows:
In Sec. \ref{Sec_2}, we discuss the relation
between the QCD thermodynamic potential and the RW periodicity.
In Sec. \ref{Sec_3}, formalism of the PNJL model is explained,
and the properties of the model in the
imaginary chemical potential region is discussed.
Sec. \ref{Sec_4} is devoted to present numerical results
calculated by the PNJL model.
The summary is given in Sec. \ref{Sec_5}.

\section{QCD PARTITION FUNCTION AND RW PERIODICITY}
\label{Sec_2}
Before going to the 2+1-flavor case,
we consider non-degenerate three-flavor QCD with
imaginary $\mu_{f}$ $(f=u,d,s)$.
For later convenience,
we introduce the dimensionless chemical potentials
$\theta_{f}$ as $\mu_{f}=i\theta_{f}T$.
In Euclidean spacetime with the time interval $\tau\in[0,\beta=1/T]$,
the QCD partition function $Z_{\rm QCD}$ is defined by
 \begin{align}
 Z_{\rm QCD}(\theta_{f})=\int\mathcal{D}A\mathcal{D}\bar{q}\mathcal{D}q
 \exp\left[-S_{\rm QCD}\right]
 \end{align}
 having the action
 \begin{eqnarray}
   S_{\rm QCD}=\int d^4 x
 \left[
 \bar{q}\left(\gamma_{\mu}D_{\mu}+\hat{m}-i\frac{\hat{\theta}}{\beta}\gamma_{4}\right)q
 +\frac{1}{4g^2}\left(F^a_{\mu\nu}\right)^2
           \right],
 \label{action_QCD}
 \end{eqnarray}
where $q=(q_{u},q_{d},q_{s})^{\rm T}$ is the quark field,
$\hat{m}=\textrm{diag}(m_{u},m_{d},m_{s})$
is the current-quark mass matrix, and $D_{\mu}=\partial_{\mu}+iA_{\mu}$ is the
covariant derivative including the gluon field
$A_{\mu}=gA^{a}_{\mu}\lambda^a/2$
with the gauge coupling $g$ and the Gell-Mann matrices $\lambda^a$.
For the quark fields, the anti-periodic boundary conditions
$q_{f}(\beta,\mathbf{x})=-q_{f}(0,\mathbf{x})$ are imposed.
The dimensionless chemical-potential matrix $\hat{\theta}$ is defined by
$\hat{\theta}=\textrm{diag}(\theta_{u},\theta_{d},\theta_{s})$.

We first redefine all the quark fields as
\begin{eqnarray}
 q_{f}\rightarrow \exp\left[i\frac{\theta_{f}}{\beta}\tau\right]q_{f}.
  \label{trans_quark_field}
\end{eqnarray}
The integral measure is unchanged under Eq.~(\ref{trans_quark_field})
and $Z_{\rm QCD}$ is transformed into
 \begin{align}
  \begin{aligned}
 &Z_{\rm QCD}(\theta_{f})=\int\mathcal{D}A\mathcal{D}\bar{q}\mathcal{D}q
 \exp\left[-S_{\rm QCD}\right],
 \\
 & S_{\rm QCD}=\int d^4 x
 \left[
 \bar{q}(\gamma_{\mu}D_{\mu}+\hat{m})q
 +\frac{1}{4g^2}\left(F^{a}_{\mu\nu}\right)^2
  \right]
   \end{aligned}
 \label{action_QCD_transform}
 \end{align}
 with the boundary conditions
 \begin{eqnarray}
  q_{f}(\beta,\mathbf{x})=-\textrm{e}^{i\theta_{f}}q_{f}(0,\mathbf{x}).
   \label{boundary_before}
 \end{eqnarray}
Now, we consider $\mathbb{Z}_{3}$ transformation
defined by
\begin{align}
 & q_{f}\rightarrow U_{k}q_{f},
 \label{quark_Z3_trans}
 \\
 & A_{\mu}\rightarrow U_{k}A_{\mu}U^{-1}_{k}+i(\partial_{\mu}U_{k})U^{-1}_{k},
 \label{gluon_Z3_trans}
 \\
 & U_{k}=\exp\left[i\frac{2\pi k}{3}\frac{\tau}{\beta}\right],\ \ k\in\mathbb{Z}.
 \label{Z3_trans}
\end{align}
The functional form of $Z_{\rm QCD}$
keeps the form of Eq.~(\ref{action_QCD_transform})
under the $\mathbb{Z}_{3}$ transformation,
but the boundary conditions are changed into
\begin{eqnarray}
 q_{f}(\beta,\mathbf{x})=
  -\exp\left[i\left(\theta_{f}-\frac{2\pi k}{3}\right)\right]q_{f}(0,\mathbf{x}).
  \label{boundary_condition_final}
\end{eqnarray}
Equations~(\ref{action_QCD_transform}), (\ref{boundary_before})
and (\ref{boundary_condition_final})
give the equality
\begin{eqnarray}
 Z_{\rm QCD}(\theta_{f}-2\pi k/3)=Z_{\rm QCD}(\theta_{f}).
\end{eqnarray}
The QCD partition function thus has
the periodicity of $2\pi/3$ in $\theta_{f}$, which is nothing but the RW periodicity.

The RW periodicity of $Z_{\rm QCD}$ can be interpreted as
the invariance under the extended $\mathbb{Z}_{3}$
transformation~\cite{Sakai_PRD77_051901, Sakai_PRD78_036001,
Sakai_PRD78_076007, Sakai_PRD79_096001},
composed of the shift $\theta_{f}\ \rightarrow \theta_{f}+2\pi k/3$
and Eqs.~(\ref{quark_Z3_trans}) - (\ref{Z3_trans}).
The QCD thermodynamic potential $\Omega_{\rm QCD}$ (per unit volume) is
related with $Z_{\rm QCD}$ as $\Omega_{\rm QCD}=-T\ln Z_{\rm QCD}$.
Therefore, $\Omega_{\rm QCD}$ also has the RW periodicity
when $Z_{\rm QCD}$ is invariant under
the extended $\mathbb{Z}_{3}$ transformation.

The discussions mentioned above can be applied
to the 2+1-flavor case by setting $\theta_{u}=\theta_{d}\equiv \theta_{l}$.
Hence, one can find that $\Omega_{\rm QCD}$ with condition (I)
has the RW periodicity because of its invariance under the
extended $\mathbb{Z}_{3}$ transformation.
Meanwhile, when any one of $\theta_{f}$ is fixed to a constant value,
for example $\theta_{s}=0$ in condition (II),
the RW periodicity disappears anymore since
one cannot make the shift
$\theta_{f}\rightarrow \theta_{f}+2\pi k/3$ for fixed $\theta_{f}$.
This is the reason why
the RW periodicity does not exist for condition (II).
In the next section, we formulate the 2+1-flavor PNJL model and
show that the PNJL model also possesses the same properties
discussed in this section.

\section{PNJL MODEL}
\label{Sec_3}
The Lagrangian of PNJL model in Euclidean spacetime
is formulated by
\begin{align}
 \mathcal{L}_{\rm PNJL}=
 &\bar{q}\left(\gamma_{\mu}D_{\mu}+\hat{m}-i\frac{\hat{\theta}}{\beta}\gamma_{4}\right)q
 +\mathcal{U}
 \notag \\
 &-G_{\rm s}\sum_{a=0}^{8}
 \left[
 (\bar{q}\lambda^a q)^2+(\bar{q}i\gamma^5\lambda^a q)^2
 \right]
 \notag \\
 &+K\left[\textrm{det}_{f}\bar{q}(1+\gamma^5)q
 +\textrm{det}_{f}\bar{q}(1-\gamma^5)q\right],
 \label{Lag_PNJL}
\end{align}
where the definitions of $q$, $\hat{m}$ and $\hat{\theta}$
are the same as in Eq.~(\ref{action_QCD}),
but the covariant derivative has the form
$D_{\mu}=\partial_{\mu}+ig\delta_{\mu 4}A^{a}_{\mu}\lambda^a/2$
in the present PNJL model.
The Polyakov-loop potential $\mathcal{U}$ is a
function of Polyakov loop $\Phi$ and its conjugate $\Phi^{\ast}$.
The definitions of these quantities are
\begin{align}
 \Phi = \frac{1}{3}{\rm Tr}_{\rm c}(L),\ \ \
 \Phi^{\ast} = \frac{1}{3}{\rm Tr}_{\rm c}({L}^{\dag}),
\end{align}
where $L= \exp[i\beta A_4]=\exp[i\beta{\rm diag}(A_4^{11},A_4^{22},A_4^{33})]$
for the classical gauge fields $A_4^{ii}$ satisfying $A_4^{11}+A_4^{22}+A_4^{33}=0$,
and the trace is taken in color space.
We use the logarithm type of
\begin{align}
 &\mathcal{U}
 =T^4\left[-\frac{a(T)}{2}\Phi\Phi^{\ast}+b(T)\ln H\right],
 \label{Polyakov_potential}
 \\
 &a(T)=a_{0}+a_{1}\left(\frac{T_{0}}{T}\right)+a_{2}\left(\frac{T_{0}}{T}\right)^2,
 \ b(T)=b_{3}\left(\frac{T_{0}}{T}\right)^3,
 \\
 &H=1-6\Phi\Phi^{\ast}+4(\Phi^3+\Phi^{\ast 3})-3(\Phi\Phi^{\ast})^2
 \label{Haar_measure}
\end{align}
in Ref.~\cite{Rossner}. Note that Eq.~(\ref{Polyakov_potential}) preserves the $\mathbb{Z}_{3}$
symmetry.

The original value of $T_{0}$ is fitted to 270 MeV
so as to reproduce the deconfinement transition temperature
in the pure gauge limit~\cite{Boyd, Kaczmarek}.
When the dynamical quarks are taken into account,
the value of $T_{0}=270$ MeV predicts higher deconfinement transition temperature
than LQCD prediction, $T_{\rm c}\sim 160$ MeV at $\theta_{f}=0$
\cite{Fodor_Katz_tem,Borsanyi, Soldner, Kanaya, Laermann}.
The calculation in Ref.~\cite{Sasaki_EPNJL}
provides lower $T_{\rm c}$ at $\theta_{f}=0$ by refitting $T_{0}$ to a lower value,
but we keep the original value to concentrate on qualitative discussions.

In the quark-quark interaction terms, $G_{\rm s}$ is the strength of the
scalar-type four-point interaction
and $K$ is the strength of the Kobayashi-Maskawa-'t Hooft (KMT) interaction
\cite{tHooft, Kobayashi-Maskawa, Kobayashi-Kondo-Maskawa}.
The determinant in the KMT interaction term is taken in flavor space.
The KMT interaction explicitly breaks $\textrm{U}_{\rm A}(1)$ symmetry
and is necessary to reproduce the measured mass of $\eta'$ meson
at vacuum.

The mean-field approximation yields the
thermodynamic potential $\Omega_{\rm PNJL}$
(per unit volume) as 
   \begin{align}
   &\Omega_{\rm PNJL}=
   2G_{\rm s}\sum_{f=u,d,s}\sigma^2_{f}
   -4K\sigma_{u}\sigma_{d}\sigma_{s}
   +\mathcal{U}
   \notag \\
   &-\frac{2}{\beta}\sum_{f=u,d,s}\int\frac{d^3\mathbf{p}}{(2\pi)^3}
   \bigl[
   3\beta E_{f}
   \notag \\
   &+
   \ln(1+3\Phi\textrm{e}^{-\beta(E_{f}-\mu_{f})}
   +3\Phi^{\ast}\textrm{e}^{-2\beta(E_{f}-\mu_{f})}
   +\textrm{e}^{-3\beta(E_{f}-\mu_{f})})
   \notag \\
   &+
   \ln(1+3\Phi^{\ast}\textrm{e}^{-\beta(E_{f}+\mu_{f})}
   +3\Phi\textrm{e}^{-2\beta(E_{f}+\mu_{f})}
   +\textrm{e}^{-3\beta(E_{f}+\mu_{f})})
   \bigr],
   \label{thermo_PNJL}
   \end{align}
 where $\mu_{f}=i\theta_{f}T$,
 $\sigma_{f}=\braket{\bar{q}_{f}q_{f}}$
 and $E_{f}=\sqrt{\mathbf{p}^2+M^2_{f}}$
 with the constituent-quark masses
  \begin{align}
   \begin{aligned}
  & M_{f}=m_{f}-4G_{\rm s}\sigma_{f}+2K\sigma_{f'}\sigma_{f''},
  \\
  & \left(f\ne f',\ \ f'\ne f'',\ \ f\ne f''\right).
   \end{aligned}
  \end{align}
 Note that
 $\theta_{u}=\theta_{d}\equiv \theta_{l}$,
 $\sigma_{u}=\sigma_{d}$
 and $E_{u}=E_{d}$ in the 2+1-flavor case.
 We introduce the three-dimensional cutoff $\Lambda$
 to regularize the vacuum term in Eq.~(\ref{thermo_PNJL}).
 The variables $X=\left\{\sigma_{l}, \sigma_{s},\Phi, \Phi^{\ast}\right\}$
 are determined by the stationary conditions,
 \begin{eqnarray}
  \frac{\partial \Omega_{\rm PNJL}}{\partial X}=0,\ \
   X=\left\{\sigma_{l}, \sigma_{s},\Phi, \Phi^{\ast}\right\}.
 \end{eqnarray}
 The parameters used in the present PNJL model
 are summarized in TABLE~\ref{table1}.

 \begin{table}[t]
\begin{center}
\caption{
 Summary of the parameter set used in the present PNJL model.
 The panels (a) and (b) are the parameter set
 in $\mathcal{U}$~\cite{Rossner} and the NJL part
 ~\cite{Reinberg_Klevansky_Hufner, Klevansky_review},
 respectively.
   }
\begin{tabular}{c|ccccc}
\hline \hline
 (a) & $a_{0}$ & $a_{1}$ & $a_{2}$ & $b_{3}$ & $T_{0}$ [MeV] \\
 & 3.51 & - 2.47 & 15.2 & - 1.75 & 270 \\ \hline
 (b) & $m_{l}$ [MeV] & $m_{s}$ [MeV] & $\Lambda$ [MeV] & $G_{\rm s}\Lambda^2$
             & $K\Lambda^5$ \\
       & 5.5 & 140.7 & 602.3 & 1.835 & 12.36 \\ \hline \hline
\end{tabular}
\label{table1}
\end{center}
 \end{table}
 
 Under the extended $\mathbb{Z}_{3}$ transformation,
 the Polyakov loop behaves as $\Phi\rightarrow \Phi\textrm{e}^{-2\pi ik/3}$
 and is not invariant.
 It is more convenient to define the
 flavor-dependent modified Polyakov loop and
 its conjugate~\cite{Sakai_JPhys} as
 \begin{eqnarray}
  \Psi_{f}=\textrm{e}^{i\theta_{f}}\Phi,\ \ \Psi^{\ast}_{f}=\textrm{e}^{-i\theta_{f}}\Phi^{\ast}.
 \end{eqnarray}
 The extended $\mathbb{Z}_{3}$ transformation
 leaves these quantities invariant.
 After rewriting Eq.~(\ref{thermo_PNJL}) by $\Psi_{f}$ and $\Psi^{\ast}_{f}$,
 we can reach the expression
   \begin{align}
   &\Omega_{\rm PNJL}=
   2G_{\rm s}\sum_{f=u,d,s}\sigma^2_{f}
   -4K\sigma_{u}\sigma_{d}\sigma_{s}
   +\mathcal{U}
   \notag \\
   &-\frac{2}{\beta}\sum_{f=u,d,s}\int\frac{d^3\mathbf{p}}{(2\pi)^3}
   \bigl[
   3\beta E_{f}
   \notag \\
   &+
   \ln(1+3\Psi_{f}\textrm{e}^{-\beta E_{f}}
   +3\Psi^{\ast}_{f}\textrm{e}^{-2\beta E_{f}}\textrm{e}^{3i\theta_{f}}
    +\textrm{e}^{-3\beta E_{f}}\textrm{e}^{3i\theta_{f}})
    \notag \\
   &+
   \ln(1+3\Psi^{\ast}_{f}\textrm{e}^{-\beta E_{f}}
   +3\Psi_{f}\textrm{e}^{-2\beta E_{f}}\textrm{e}^{-3i\theta_{f}}
   +\textrm{e}^{-3\beta E_{f}}\textrm{e}^{-3i\theta_{f}})
   \bigr].
   \label{thermo_PNJL2}
   \end{align}
   The $\theta_{f}$-dependence of Eq.~(\ref{thermo_PNJL2}) is
   embedded in the extended $\mathbb{Z}_{3}$ symmetric quantities
   $\{\textrm{e}^{\pm 3i\theta_{f}},\Psi_{f},\Psi^{\ast}_{f}\}$.
   Obviously, $\Omega_{\rm PNJL}$ is invariant
   under the extended $\mathbb{Z}_{3}$ transformation
   and hence $\Omega_{\rm PNJL}$ has the RW periodicity in general.
   Once any one of $\theta_{f}$ is fixed to some constant value,
   however,
   the extended $\mathbb{Z}_{3}$ transformation
   changes $\Psi_{f}$ into $\Psi_{f}\textrm{e}^{-2\pi ik/3}$
   and thereby $\Omega_{\rm PNJL}$ does not become invariant.
   It is thus concluded that
   $\Omega_{\rm PNJL}$ has the same properties as $\Omega_{\rm QCD}$
   on the RW periodicity.

   \section{NUMERICAL RESULTS}
   \label{Sec_4}
   We show numerical results calculated by the PNJL model.
   In calculations of thermodynamic quantities
   and the QCD phase diagram,
   both conditions (I) and (II) are considered.
   We pick up $\Omega_{\rm PNJL}$ and the quark number density $n_{q}$
   as the thermodynamic quantities,
   and calculate $\theta_{l}$-dependence for $T=200, 250$ MeV.
   In the results of condition (I), the RW periodicity can be seen.
   On the contrary, there is no RW periodicity for condition (II),
   as expected in Sec.~\ref{Sec_3}.
   In the QCD phase diagram,
   we find for condition (II)
   that the crossover chiral transition line is discontinuous
   at some value of $\theta_{l}$.
   In addition, the first-order phase transition line appears
   as is the RW phase transition line,
   and can be fitted by a polynomial function of $\theta_{l}$.
   Finally, the up- and strange-quark number densities are calculated
   under the situation that no RW periodicity exists.
   We show that the non-analyticity in the number densities disappears
   below some constant value of $\theta_{l}$ or $\theta_{s}$.

     \begin{figure}
  \begin{center}
   \includegraphics[width=0.45\textwidth]{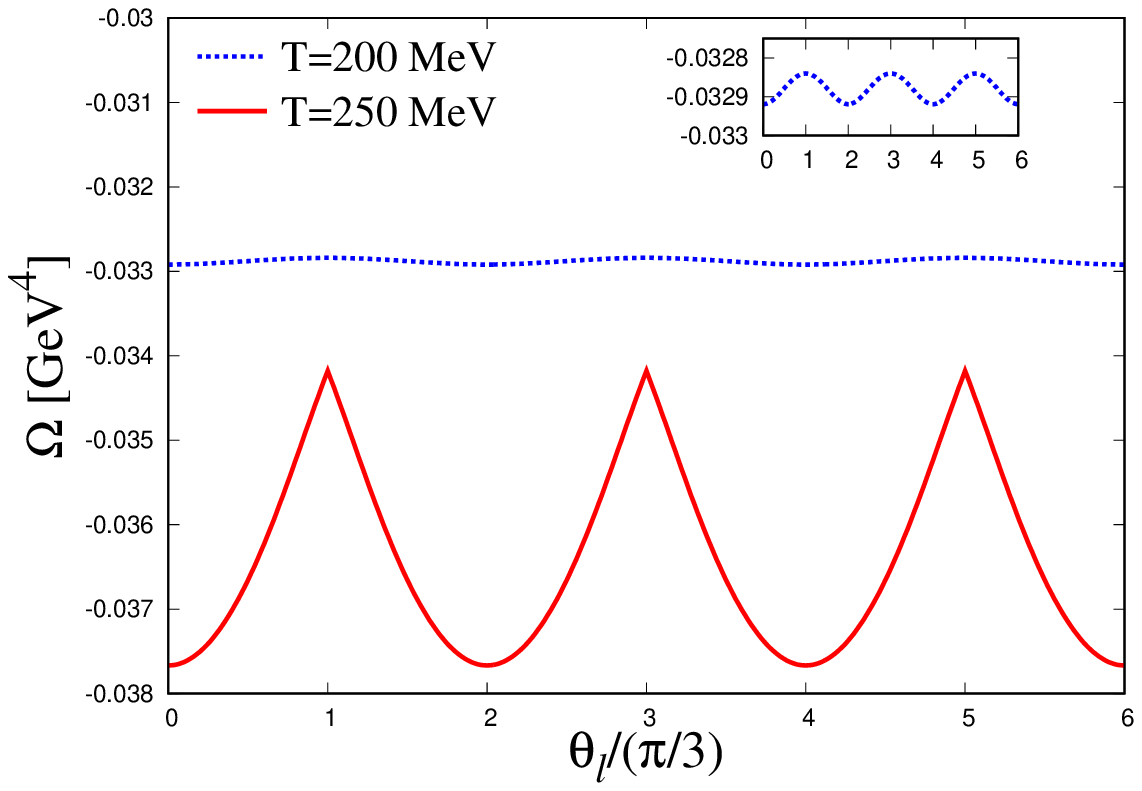}
   \includegraphics[width=0.45\textwidth]{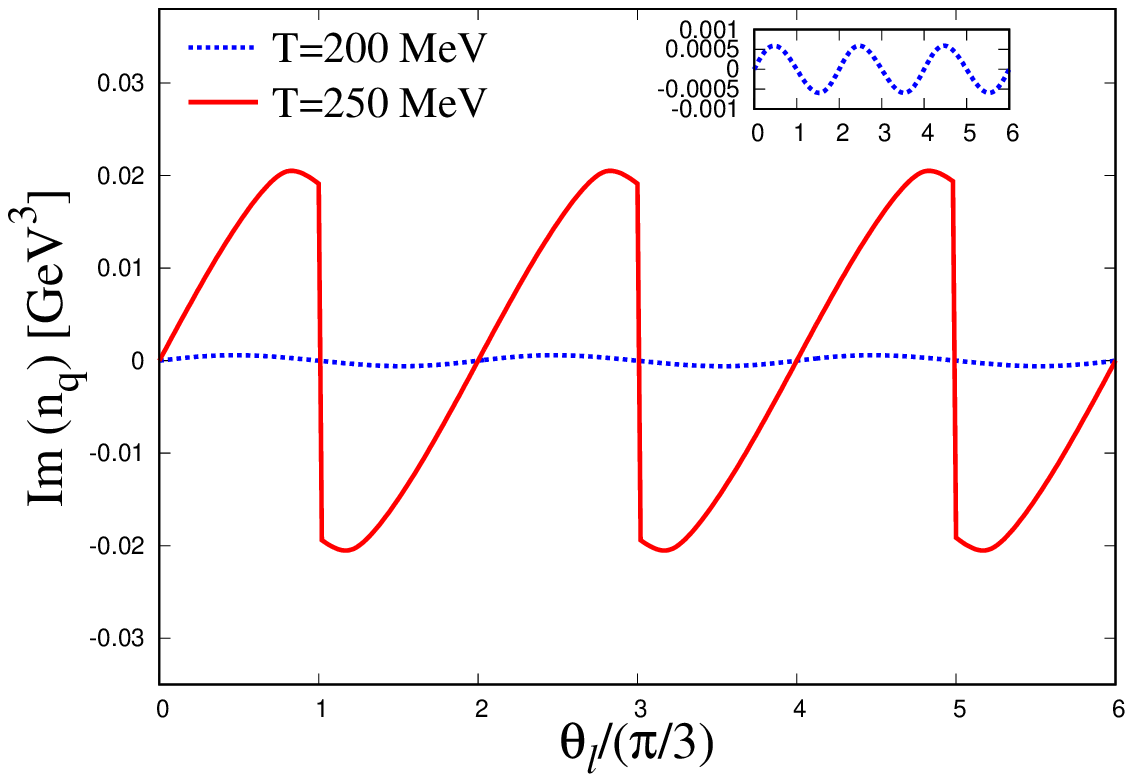}
   \end{center}
\caption{
     The $\theta_{l}$-dependence of $\Omega_{\rm PNJL}$
      and the imaginary part of the quark number density
      $\textrm{Im}(n_{q})$ calculated by the PNJL model
      for condition (I).
      The solid line is the results for $T=250$ MeV and the dotted line for $T=200$ MeV.
}
\label{thermodynamic_quantity1}
     \end{figure}

     \begin{figure}
  \begin{center}
   \includegraphics[width=0.45\textwidth]{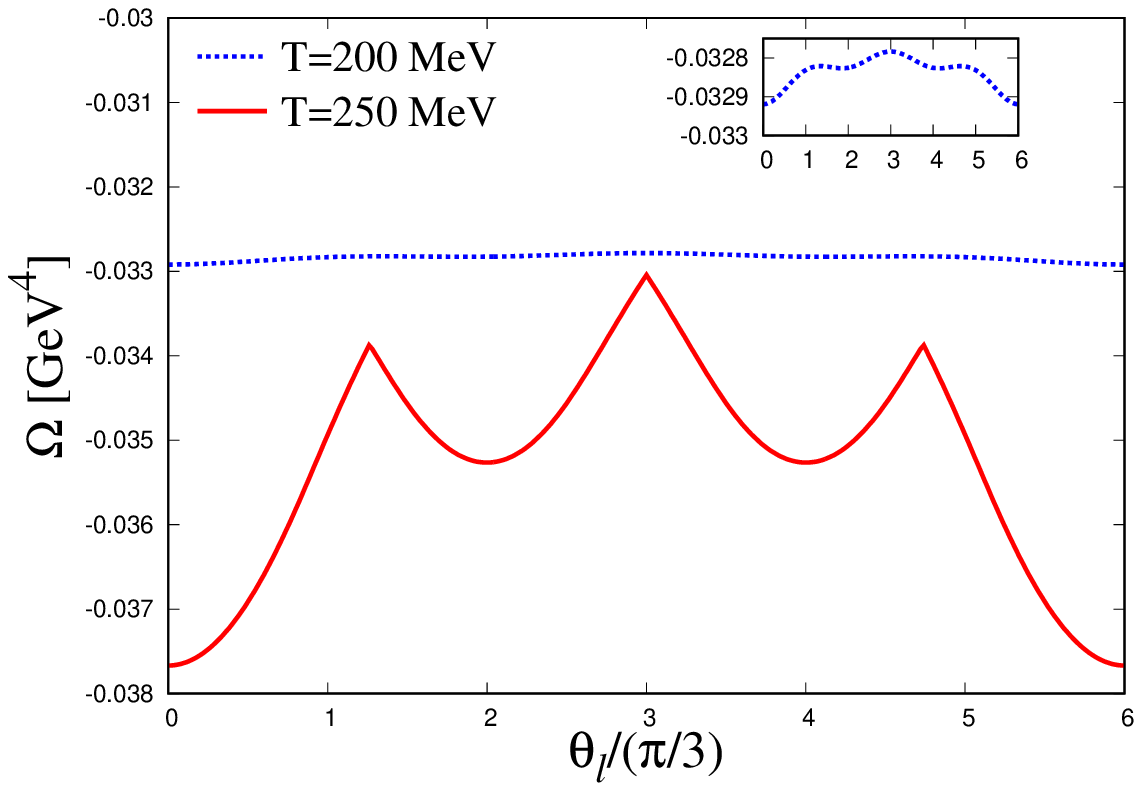}
   \includegraphics[width=0.45\textwidth]{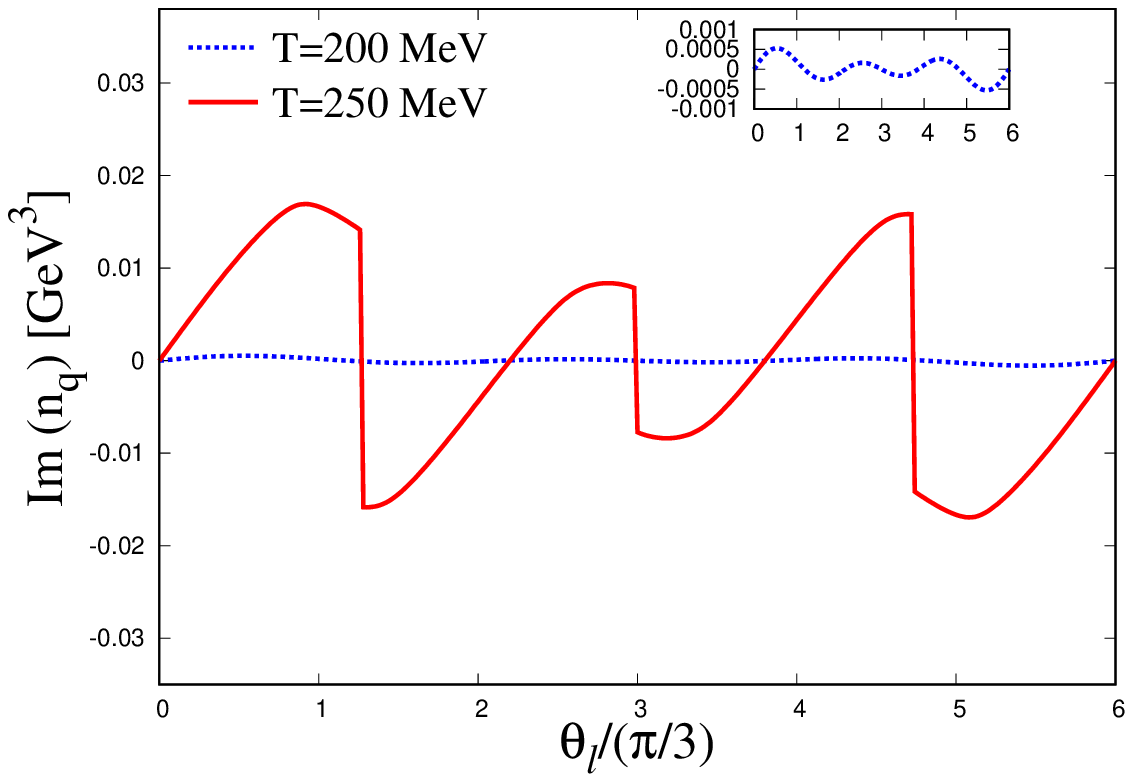}
   \end{center}
\caption{
     The $\theta_{l}$-dependence of $\Omega_{\rm PNJL}$
      and the imaginary part of the quark number density $\textrm{Im} (n_{q})$
      calculated by the PNJL model for condition (II).
     The meanings of each line are same as in Fig.~\ref{thermodynamic_quantity1}.
}
\label{thermodynamic_quantity2}
     \end{figure}

    \subsection{BEHAVIOR OF THERMODYNAMIC QUANTITIES}
    The quark number density $n_{q}$
    is obtained by the relation
    \begin{eqnarray}
     n_{q}=\sum_{f=u,d,s}n_{f}
      =i\beta\sum_{f=u,d,s}\frac{\partial}{\partial\theta_{f}}\Omega_{\rm PNJL},
      \label{number_density}
    \end{eqnarray}
    where $n_{f}$ is the number density of the quark with flavor $f$.
    Using Eq.~(\ref{number_density}), we can see that
    the condition to exist the RW periodicity in $n_{q}$
    is equivalent to that in $\Omega_{\rm PNJL}$.
    Since $\Omega_{\rm PNJL}$ is charge-even,
    $n_{f}$ is charge-odd; Namely,
    $\Omega_{\rm PNJL}(\theta_{f})=\Omega_{\rm PNJL}(-\theta_{f})$
    and
    $n_{q}(-\theta_{f})=-n_{q}(\theta_{f})$.

    Figure~\ref{thermodynamic_quantity1} presents
    $\Omega_{\rm PNJL}$ and the imaginary part 
    of $n_{q}$, $\textrm{Im}(n_{q})$,
    for condition (I),
    as a function of $\theta_{l}$.
    The dotted line denotes the results for $T=200$ MeV
    and the solid line does for $T=250$ MeV.
    Both $\Omega_{\rm PNJL}$
    and $n_{q}$ have the RW periodicity and
    are smooth for any $\theta_{l}$ when $T=200$ MeV.
    Meanwhile, $\Omega_{\rm PNJL}$ has cusps
    at $\theta_{l}=\pi/3$ mod $2\pi/3$ and $n_{q}$ becomes discontinuous there
    for $T=250$ MeV.
    These singularities mean the first-order RW phase transition,
    and indicate that the RW endpoint is located in $200< T < 250$ MeV
    (see Fig~\ref{Fig_phase_diagram1}).

    Now, we concentrate on the region of $0\le \theta_{l}\le 2\pi/3$.
    For charge-even quantities $\mathcal{O}_{\rm even}$ with the RW periodicity,
    such as $\Omega_{\rm PNJL}$,
    the relation
    \begin{align}
     \mathcal{O}_{\rm even}(\theta_{l}-\epsilon)
     &=\mathcal{O}_{\rm even}(-\theta_{l}+\epsilon)
     \notag \\
     &=\mathcal{O}_{\rm even}(-\theta_{l}+2\pi/3+\epsilon)
    \end{align}
    is obtained,
    where $\epsilon$ is a positive infinitesimal quantity.
    If the gradient
    \begin{eqnarray}
     \lim_{\theta_{l}\rightarrow \pi/3\pm 0}\frac{d\mathcal{O}_{\rm even}}{d\theta_{l}}
    \end{eqnarray}
    is neither zero nor infinity,
    charge-even quantities have a cusp at $\pi/3$.
    On the other hand, charge-odd quantities $\mathcal{O}_{\rm odd}$
    possessing the RW periodicity,
    such as $\textrm{Im} (n_{q})$,
    satisfy
    \begin{align}
     \mathcal{O}_{\rm odd}(\theta_{l}-\epsilon)&=
     -\mathcal{O}_{\rm odd}(-\theta_{l}+\epsilon)
     \notag \\
     &=-\mathcal{O}_{\rm odd}(-\theta_{l}+2\pi/3+\epsilon).
    \end{align}
    Hence, discontinuity is seen at $\theta_{l}=\pi/3$
    for charge-odd quantities in high $T$ region
    \cite{Sakai_PRD77_051901, Sakai_PRD78_036001,
    Sakai_PRD78_076007, Kouno_JPhys},
    where
    \begin{eqnarray}
     \lim_{\theta_{l}\rightarrow \pi/3\pm 0}\mathcal{O}_{\rm odd}(\theta_{l})\ne 0.
    \end{eqnarray}
    Due to these singularities, the analytic continuation
    from the imaginary $\mu_{l}$ to the real one
    is limited up to $\theta_{l}=\pi/3$,
    particularly for the high $T$ region.

    Figure~\ref{thermodynamic_quantity2}
    is same as Fig.~\ref{thermodynamic_quantity1},
    but for condition (II).
    It is clearly seen that the RW periodicity is lost,
    but $\theta_{l}$-dependence
    is similar to each other
    between Figs.~\ref{thermodynamic_quantity1} and
    \ref{thermodynamic_quantity2}.
    In particular, the first-order phase transition
    still takes place for $T=250$ MeV,
    and it is expected that its endpoint is located in $200<T<250$ MeV
    (see Fig.~\ref{Fig_phase_diagram2}).
    We refer to this transition as the first-order ``RW-like phase transition''.
    It should be noted that
    the RW-like phase transition occurs at $\theta_{l}=0.42\pi$
    for $T=250$ MeV.
    This result indicates that the
    region needed to the analytic continuation becomes broader
    for condition (II) than (I), as
    already pointed out in Ref.~\cite{Bonati}.

    \subsection{PHASE DIAGRAM}
  To determine the crossover chiral and
  deconfinement transition lines,
  we calculate the
  pseudo-critical temperature of each
  transition by the peak position of susceptibilities
  for given $\theta_{l}$.
  According to Ref.~\cite{Sasaki_sus},
  the susceptibilities $\chi_{ij}$ of $\{\sigma_{l},\sigma_{s},\Phi,\Phi^{\ast}\}$
  can be calculated by the inverse of dimensionless
  curvature matrix,
  $\chi_{ij}=\left(\mathcal{C}^{-1}\right)_{ij}$,
  where
 \begin{eqnarray}
\mathcal{C}=
\begin{pmatrix}
T^2c_{\sigma_{l}\sigma_{l}}
&
T^2c_{\sigma_{l}\sigma_{s}}
&
T^{-1}c_{\sigma_{l}\Phi}
&
T^{-1}c_{\sigma_{l}\Phi^{\ast}}
\\
T^2c_{\sigma_{s}\sigma_{l}}
&
T^2c_{\sigma_{s}\sigma_{s}}
&
T^{-1}c_{\sigma_{s}\Phi}
&
T^{-1}c_{\sigma_{s}\Phi^{\ast}}
\\
T^{-1}c_{\Phi\sigma_{l}}
&
T^{-1}c_{\Phi\sigma_{s}}
&
T^{-4}c_{\Phi\Phi}
&
T^{-4}c_{\Phi\Phi^{\ast}}
\\
T^{-1}c_{\Phi^{\ast}\sigma_{l}}
&
T^{-1}c_{\Phi^{\ast}\sigma_{s}}
&
T^{-4}c_{\Phi^{\ast}\Phi}
&
T^{-4}c_{\Phi^{\ast}\Phi^{\ast}}
\end{pmatrix}
\label{curvature_matrix}
 \end{eqnarray}
 with the abbreviation of
 \begin{eqnarray}
  c_{xy}=\frac{\partial^2\Omega_{\rm PNJL}}{\partial x\partial y},\ \ x,y=
   \left\{\sigma_{l}, \sigma_{s}, \Phi, \Phi^{\ast}\right\}.
 \end{eqnarray}

      \begin{figure}[b]
   \begin{center}
    \includegraphics[width=0.45\textwidth]{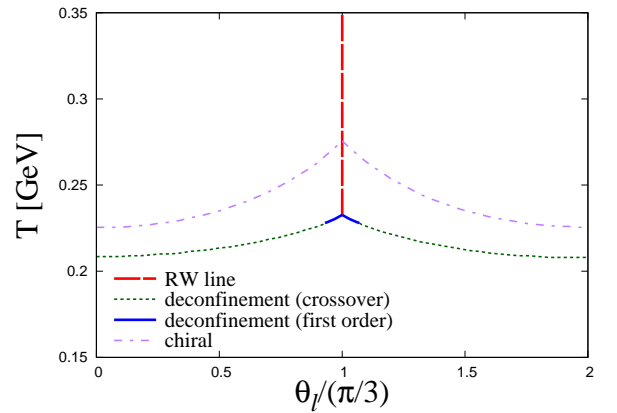}
   \end{center}
      \caption{
       The phase diagram in the $\theta_{l}$-$T$ plane
       for condition (I).
       The dashed line means the RW phase transition line.
       The crossover (first-order) deconfinement transition line
       is represented by dotted (solid) line.
       The dot-dashed line corresponds to the crossover chiral transition line.
      }
      \label{Fig_phase_diagram1}
      \end{figure}

  At the RW or RW-like phase transition points,
  $n_{q}$ becomes discontinuous,
  as already shown in Figs.~\ref{thermodynamic_quantity1}
  and~\ref{thermodynamic_quantity2}.
  This singular behavior is a good indicator to determine the
  location of the RW or RW-like phase transition points~\cite{Kouno_JPhys},
  and we use this property to determine the RW
  or RW-like phase transition lines.
  The usefulness of $n_{q}$ to search the RW phase transition point
  is also discussed from the view point of topological order~\cite{Kashiwa_holonomy}.

  Figure~\ref{Fig_phase_diagram1} presents
  the QCD phase diagram in the $\theta_{l}$-$T$ plane
  for condition (I).
  We only consider the region $\theta_{l}\in[0,2\pi/3]$
  because of the RW periodicity.
  The dot-dashed line is the crossover chiral transition line,
  and the dotted line is the deconfinement one.
  The solid line denotes the first-order deconfinement transition line,
  connected to the endpoint of the RW transition line
  represented by the dashed line.
  The RW endpoint is located at
  $(T^{\rm RW},\theta^{\rm RW}_{l})=(0.233\ \textrm{GeV}, \pi/3)$. 
  The chiral transition is crossover in the entire region,
  while the deconfinement transition becomes first-order,
  which means that the RW endpoint is a triple point.
  
  We comment on the order of the RW endpoint.
  The order of deconfinement transition depends on
  the Polyakov-loop potential $\mathcal{U}$
  taken~\cite{Kouno_JPhys, Sakai_JPhys} and
  the entanglement coupling $G_{\rm s}(\Phi,\bar{\Phi})$
  \cite{Sasaki_EPNJL,Sakai_EPNJL,Sakai_Sasaki_JPhys}.
  For example, the deconfinement transition becomes
  second order \cite{Kouno_JPhys, Sakai_JPhys},
  if we choose
  \begin{eqnarray}
   \mathcal{U}=-bT\left[
                   54\textrm{e}^{-aT}\Phi\Phi^{\ast}+
                   \log H
                  \right]
  \end{eqnarray}
  as a form of $\mathcal{U}$
  \cite{Fukushima1}, where $H$ is defined in Eq.~(\ref{Haar_measure})
  and $a, b$ are parameters.
  In this case, the RW endpoint becomes a
  tri-critical point.
  Also in the PNJL model with the entanglement coupling
  \begin{eqnarray}
   G_{\rm s}(\Phi,\Phi^{\ast})=G_{\rm s}
    \left(1-\alpha_{1}\Phi\Phi^{\ast}-\alpha_{2}(\Phi^3
     +\Phi^{\ast 3})\right)
  \end{eqnarray}
  and $(\alpha_{1}, \alpha_{2})=(0.25, 0.1)$,
  the RW endpoint becomes a tri-critical point~\cite{Sasaki_EPNJL}.
  This situation requires more robust studies to determine the order of
  the RW endpoint.

      \begin{figure}[t]
   \begin{center}
    \includegraphics[width=0.45\textwidth]{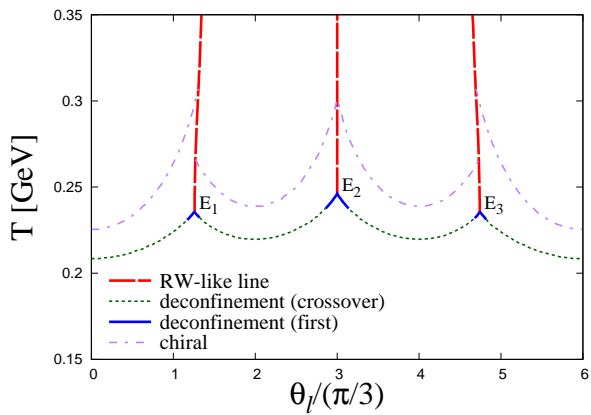}
   \end{center}
      \caption{
     The phase diagram in the $\theta_{l}$-$T$ plane
     for condition (II).
     The meanings of lines are same as in Fig.~\ref{Fig_phase_diagram2},
       except that the dashed-line denotes the RW-like phase transition line.
       Each point of $\textrm{E}_{1}$,
       $\textrm{E}_{2}$, $\textrm{E}_{3}$ stands for the triple point of the RW-like phase
       transition line.
      }
      \label{Fig_phase_diagram2}
      \end{figure}

       \begin{table}[b]
        \begin{center}
         \caption{
         The location of points $\textrm{E}_{1}, \textrm{E}_{2}$ and $\textrm{E}_{3}$
         in Fig.~\ref{Fig_phase_diagram2}.
         }
         \begin{tabular}{c|ccc}
          \hline \hline
          point & $\textrm{E}_{1}$ & $\textrm{E}_{2}$ & $\textrm{E}_{3}$ \\
          $(T,\theta_{l})$ & (0.236 GeV, 0.42$\pi$) & (0.246 GeV, $\pi$) &
                      (0.236 GeV, 1.58$\pi$)
                  \\ \hline \hline
         \end{tabular}
         \label{table2}
        \end{center}
       \end{table}

  Figure~\ref{Fig_phase_diagram2} is the
  phase diagram for condition (II).
  The meaning of lines is the same as in Fig.~\ref{Fig_phase_diagram1},
  except that the dashed line denotes the RW-like phase transition line.
  The location of points $\textrm{E}_{1}$, $\textrm{E}_{2}$
  and $\textrm{E}_{3}$ is listed in TABLE~\ref{table2}.
  The LQCD calculation of Refs.~\cite{Bonati} predicts that
  the RW-like phase transition occurs
  at $\theta_{l}\cong 0.45\pi$ for $T=208$ MeV.
  The PNJL model result $\theta_{l}=0.42\pi$
  for $\textrm{E}_{1}$ is consistent with the LQCD value
  $\theta_{l}\cong 0.45\pi$.
  
  It is found that the RW periodicity is lost, but
  the phase diagram is line symmetrical with respect to $\theta_{l}=\pi$,
  because of charge conjugation (C) symmetry of the PNJL model.
  The symmetry ensures that
  the chiral transition line has a cusp at point $\textrm{E}_{2}$.
  Meanwhile, the chiral transition line becomes discontinuous
  when it hits the RW-like line starting from points $\textrm{E}_{1}$
  and $\textrm{E}_{3}$.
  As for the first-order deconfinement line,
  it becomes symmetric due to C symmetry around point $\textrm{E}_{2}$,
  but asymmetric around points $\textrm{E}_{1}$ and $\textrm{E}_{3}$.

  In the region $\theta_{l}\in\left[0,2\pi/3\right]$,
  the RW-like phase transition starts at $\textrm{E}_{1}$,
  i.e., $(T^{\rm RW'}, \theta^{\rm RW'}_{l})
  =(0.236\ \textrm{GeV}, 0.42\pi)$.
  We fit the transition line
  by the polynomial function
  \begin{align}
   \theta_{l}(n_{\rm max})
   &=0.42\pi+\sum_{n=1}^{n_{\rm max}}a_{n}\xi^n,\ \ 
   \xi=\frac{T-T^{\rm RW'}}{T^{\rm RW'}}.
   \label{polynomial}
  \end{align}
  The transition line
  is well approximated by $\theta_{l}(n_{\rm max}=3)$
  with $a_{1}=-0.023$, $a_{2}=0.93$ and $a_{3}=-1.05$.
  The smallness of $a_{1}$ means that
  the line is nearly vertical
  in the vicinity of $\textrm{E}_{1}$ just as the RW phase transition line,
  but the transition line deviates from the vertical line
  as $T$ increases.

  The RW-like phase transition line also appears
  when we consider the imaginary isospin chemical potential
  $\mu_{I}=i\theta_{I}T$, where $\theta_{I}$ is a
  dimensionless isospin chemical potential.
  In the $\theta_{I}$-$T$ plane,
  the RW-like phase transition line is almost vertical
  and described by $\theta_{I}=\pi/2-\delta(T)$ with~\cite{Cea}
  \begin{eqnarray}
   \delta(T)=0.00016\times (T-250).
  \end{eqnarray}
  For the details, see Ref.~\cite{Sakai_JPhys, Cea}.

In Figs.~\ref{Fig_phase_diagram1} and \ref{Fig_phase_diagram2}, 
the deconfinement transition line joins the RW or RW-like endpoints, 
and the chiral transition line is higher than the deconfinement one. 
In LQCD calculation of Ref.~\cite{Bonati3}, however, 
the chiral transition line is connected to the endpoints.
At the present stage, our model cannot explain the LQCD result. 
What happens at the endpoints? This is an interesting future work 
from the theoretical point of view. 

  Finally, we compare the chiral transition line, $T=T_{\rm c}(\theta_{l})$, 
  calculated by the PNJL model with that by LQCD simulations of 
  Ref.~\cite{Bonati2}; note that $\theta_{l}$ varies with $\theta_{s}$ fixed 
at either 0 or $\theta_{l}$. 
 The ratio $R=T_{\rm c}(\theta_{l})/T_{\rm c}(0)$ is charge-even,
 and can be parametrized by~\cite{Bonati,Bonati2} 
  \begin{eqnarray}
   R=1+9\kappa\theta^2_{l}+b\theta^4_{l}
  \end{eqnarray}
  with the curvature $\kappa$ of the transition line and some constant $b$, 
  when $\theta_{l}$ is not large. 

\begin{figure}[t]
  \begin{center}
   \includegraphics[width=0.45\textwidth]{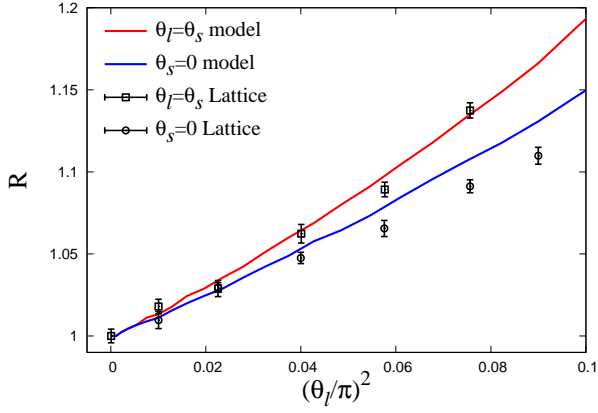}
   \end{center}
      \caption{
      $\theta^2_{l}$-dependence of ratio
      $R=T_{\rm c}(\theta_{l})/T_{\rm c}(0)$.
      The horizontal axis is normalized by $\pi^2$.
      The model calculations are represented by solid lines,
      and data with error bar mean LQCD results~\cite{Bonati2}.
}
\label{chiral_line}
     \end{figure}

  Figure~\ref{chiral_line} represents $\theta^2_{l}$-dependence
  of $R$ calculated from the PNJL model and LQCD simulations.
  The PNJL model well reproduces LQCD data for $\theta_{s}=\theta_{l}$ 
  and is almost consistent with LQCD data for $\theta_{s}=0$. 
  Thus, the present PNJL model may be good enough for qualitative analyses.   

  \subsection{ANALYTICITY OF NUMBER DENSITY}
  We calculate the imaginary part of
  up- and strange-quark number densities,
  $\textrm{Im}\left(n_{u}\right)$ and $\textrm{Im}\left(n_{s}\right)$,
  by using the PNJL model.
  We consider the situation that the RW periodicity does not exist,
  that is, some chemical potentials are fixed to constant values.
  Only in calculations of
  $\textrm{Im}\left(n_{u}\right)$,
  $\theta_{d}$ and $\theta_{s}$ are treated as constants.
  As for calculations in
  $\theta_{s}$-dependence of $\textrm{Im}\left(n_{s}\right)$,
  we again consider $\theta_{u}=\theta_{d}=\theta_{l}$
  and these are fixed to constant values.

  \begin{figure}[t]
   \begin{center}
    \includegraphics[width=0.45\textwidth]{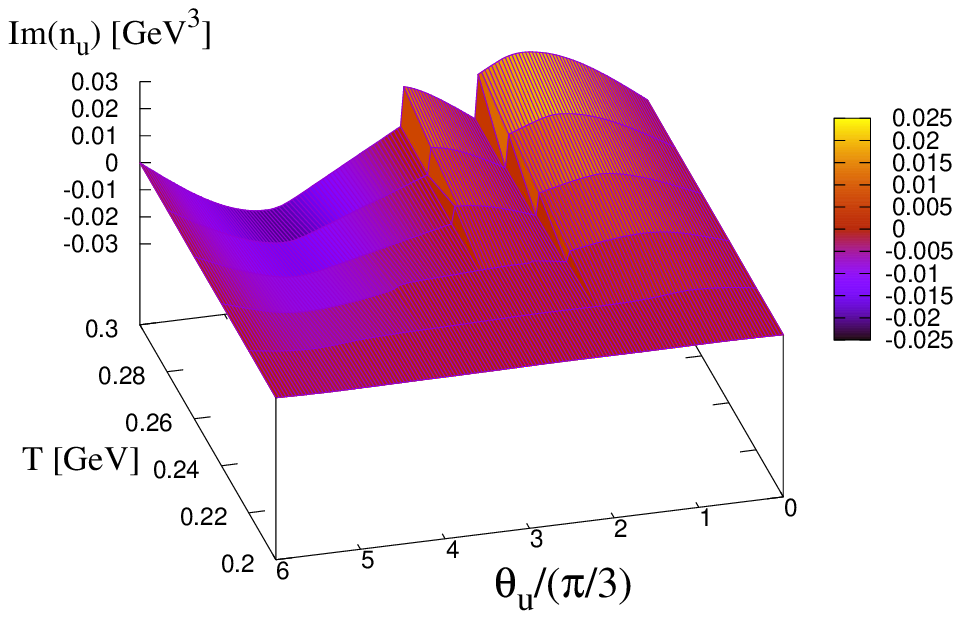}
    \includegraphics[width=0.45\textwidth]{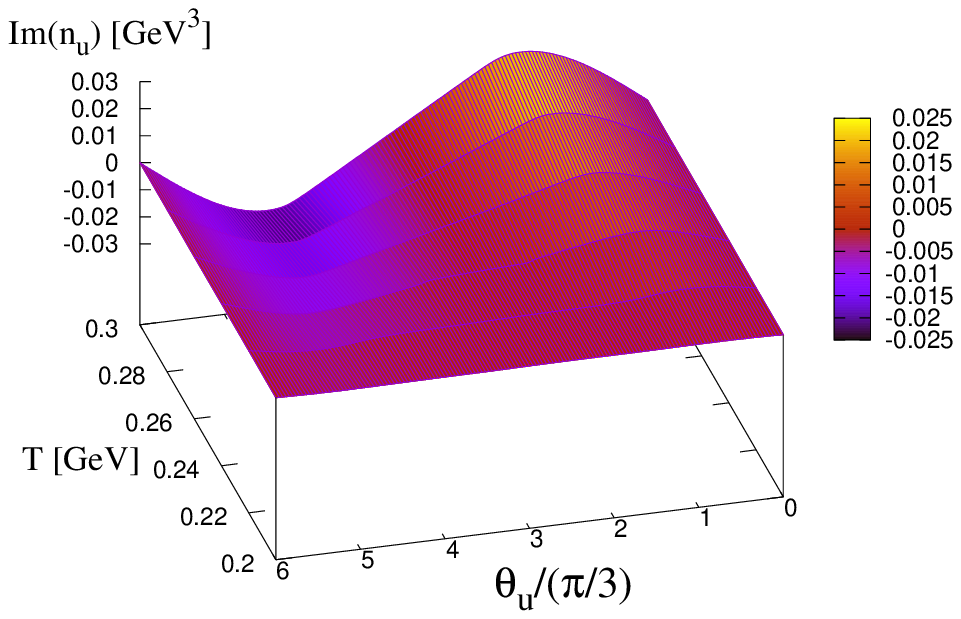}
   \end{center}
       \caption{
   The $\theta_{u}$- and $T$-dependence of $\textrm{Im}(n_{u})$.
   The upper panel is the result with
   $(\theta_{d},\theta_{s})=(\pi/4,0)$,
   and lower panel is the one with $(\theta_{d},\theta_{s})=(\pi/8,0)$.
      }
      \label{Fig_nu_3d}
  \end{figure}

  Figure~\ref{Fig_nu_3d}
  shows $\theta_{u}$- and $T$- dependence of $\textrm{Im}(n_{u})$.
  The upper panel is the result for
  $(\theta_{d},\theta_{s})=(\pi/4,0)$
  and the lower one is for
  $(\theta_{d},\theta_{s})=(\pi/8,0)$.
  In the upper panel, $\textrm{Im}(n_{u})$
  becomes discontinuous
  because of the RW-like phase transition,
  but smooth at any $T$ in the lower panel.
  We numerically checked that
  $\textrm{Im}(n_{u})$ becomes smooth
  at any $T$ when $\theta_{s}=0$ and $\theta_{l}\le \pi/8$.

  \begin{figure}
   \begin{center}
    \includegraphics[width=0.45\textwidth]{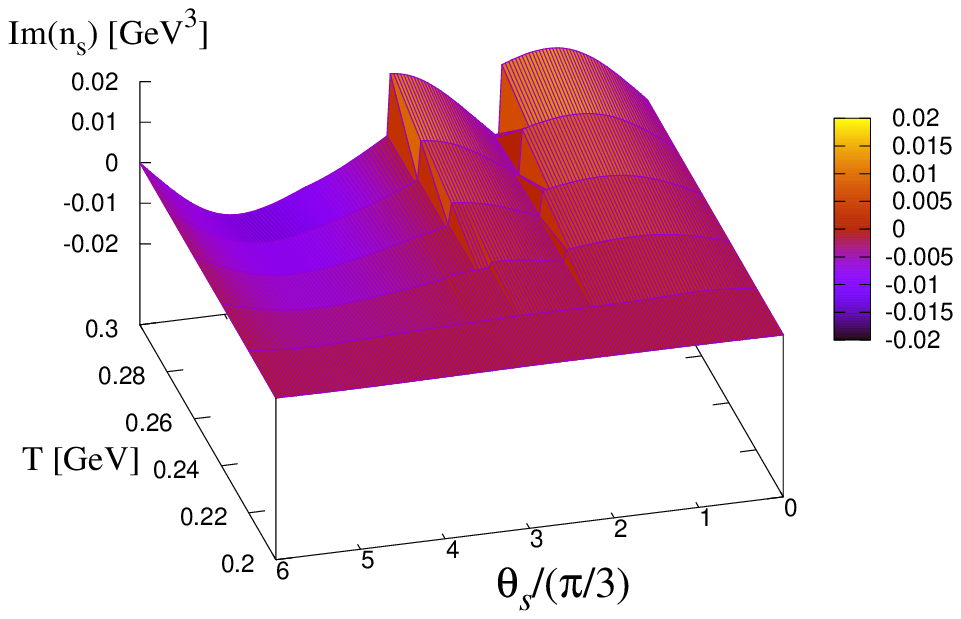}
    \includegraphics[width=0.45\textwidth]{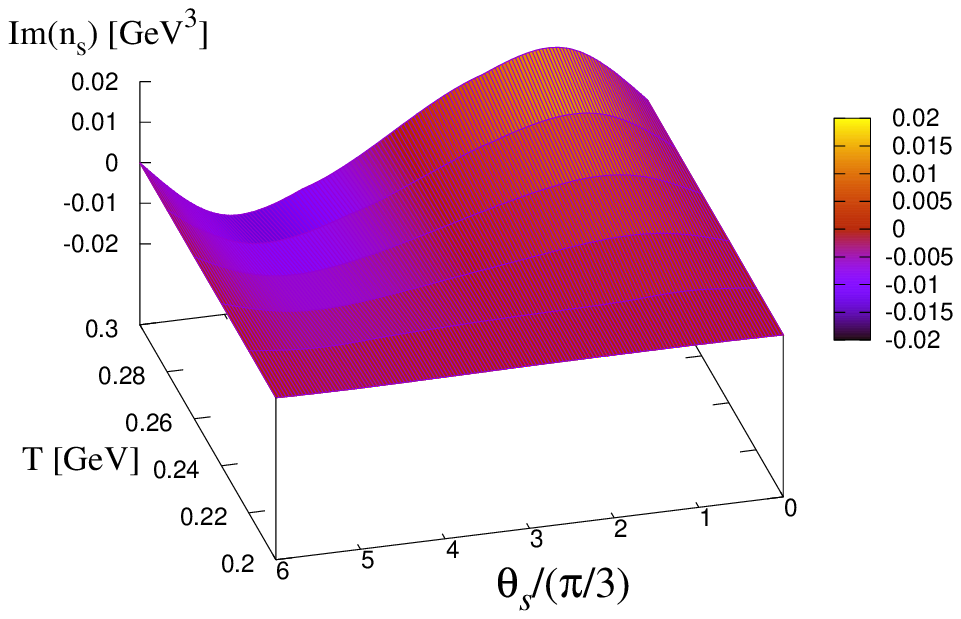}
   \end{center}
       \caption{
   The $\theta_{s}$- and $T$-dependence of $\textrm{Im}(n_{s})$.
   The upper panel is the result with
   $\theta_{l}=\pi/4$, and the lower panel is the one with $\theta_{l}=\pi/5$.
      }
      \label{Fig_ns_3d}
  \end{figure}

  Figure~\ref{Fig_ns_3d}
  is the result of $\textrm{Im}\left(n_{s}\right)$
  as a function of $\theta_{s}$ and $T$.
  The upper panel corresponds to the result
  for $\theta_{l}=\pi/4$,
  and the lower panel is the result for $\theta_{l}=\pi/5$.
  It is found that the discontinuity of $\textrm{Im}(n_{s})$
  disappears for any $T$ when $\theta_{l}=\pi/5$,
  while $\textrm{Im}(n_{s})$ becomes discontinuous
  when $\theta_{l}=\pi/4$,
  due to the RW-like phase transition.
  We also numerically confirmed that
  $\textrm{Im}(n_{s})$ has no discontinuity
  for any $T$ when $\theta_{l}\le \pi/5$.
  The results in Fig.~\ref{Fig_nu_3d} (Fig.~\ref{Fig_ns_3d})
  indicate that $n_{u}$ ($n_{s}$)
  in the real $\mu_{u}$ ($\mu_{s}$) region
  can be obtained by the analytic continuation
  from the imaginary region for any $T$.
  The present case is thus more informative
  compared to the case where the RW periodicity exists.

  The $n_{\rm s}$ in the high $T$ region
  plays a key role in determining the strength of
  the repulsive interaction, 
  \begin{eqnarray}
   \mathcal{L}_{\rm v,s}=-G_{\rm v, s}\left(\bar{s}\gamma^{\mu}s\right)^2, 
    \label{repulsive_strange}
  \end{eqnarray}
where $s$ is the strange-quark field and $G_{\rm v,s}$ is its 
strength.
  The behavior of $n_{\rm s}$ is sensitive to the value of
  $G_{\rm v,s}$, 
  because $n_{\rm s}$ is a function of
  \begin{eqnarray}
   \tilde{\mu}_{\rm s}=\mu_{\rm s}-2G_{\rm v,s}n_{\rm s}
  \end{eqnarray}
  after the mean-field approximation.

  In our previous works~\cite{Sugano},
  it was shown that
  the strength $G_{\rm v}$ of the vector-type four-quark interaction
   \begin{eqnarray}
    \mathcal{L}_{\rm v}=-G_{\rm v}(\bar{q}\gamma^{\mu}q)^2
     \label{vector_four}
   \end{eqnarray}
   can be determined from LQCD data on the quark number density $n_{q}$
   in the high $T$ region~\cite{Takahashi2, Ejiri}.
   We then pinned down the value of $G_{\rm v}$ from LQCD data on 
   $n_{q}$. However, this analysis did not consider strange quark. 
   Figure~\ref{Fig_ns_3d} indicates that 
   the analytic continuation from imaginary $\mu_s$ to real $\mu_s$ 
   works well even in the high $T$ region.
Thus, one can get reliable $n_{\rm s}$ in both the 
real- and the imaginary-$\mu_s$ region. This allows us to determine
the value of $G_{\rm v,s}$ sharply from the LQCD data. 

  The interaction described by Eq.~(\ref{repulsive_strange})
  corresponds to the interaction
  mediated by $\phi$-meson in the context of the relativistic mean
  field theory~\cite{RMF_Glendenning, RMF_Schaffner,
  RMF_Ishizuka, RMF_Tsubakihara, RMF_Weissenborn},
  and affects the maximum mass of neutron star, when 
  the strange quark exists in the inner core of neutron star.
  It is an interesting future work to investigate the interplay 
  between the $G_{\rm v,s}$ 
  determined from the LQCD data and the maximum mass, and to discuss 
  what happens in the two-solar-mass neutron star~\cite{Demorest, Antoniadis}.

  \section{SUMMARY}
  \label{Sec_5}
  In this paper, we investigated
  properties of the 2+1-flavor QCD in
  the imaginary chemical potential region with finite 
  $\mu_{l}=i\theta_{l}T$ and $\mu_{s}=i\theta_{s}T$, 
  using two approaches. One is a theoretical approach based 
  on the QCD partition function, 
  and the other is a qualitative one based on an effective model. 
  In the theoretical approach, we proved that the QCD thermodynamic 
potential $\Omega_{\rm QCD}$
  exhibits the Roberge-Weiss (RW) periodicity
  only when $\Omega_{\rm QCD}$ is invariant under the 
  extended $\mathbb{Z}_{3}$ transformation. 
  In other words, the RW periodicity disappears when
 two chemical potentials are fixed to a constant value.
  Next, we showed that the thermodynamic potential 
of the Polyakov-loop extended Nambu--Jona-Lasinio (PNJL) model 
also possesses the extended $\mathbb{Z}_{3}$ symmetry. 
We then took the PNJL model as a useful effective model. 

  Taking the PNJL model,
  we calculated $\Omega_{\rm PNJL}$, $\textrm{Im}\left(n_{q}\right)$ 
  (the imaginary part of quark number density), and the QCD phase diagram 
  as a function of $\theta_{l}$
  for two conditions; (I) $\theta_{s}=\theta_{l}$
  and (II) $\theta_{s}=0$. 
  For condition (I),
  the RW periodicity is seen in all the results.
  The structure of the phase diagram
  is similar to the one in 2-flavor case
  \cite{Sakai_JPhys}.
  As for condition (II),
  there is no RW periodicity,
  but we found that the region available for the analytic
  continuation is broader than condition (I).
  The noteworthy points on the phase diagram are the following: 
\begin{enumerate}
 \item  The crossover chiral transition line becomes
            discontinuous on the RW-like phase transition line,
 \item  The first-order deconfinement transition line
            is asymmetric with respect to the RW-like phase transition line,
            except for at $\theta_{l}=\pi$. 
 \item  The first-order RW-like phase transition line
            can be well fitted by a polynomial function of Eq.~(\ref{polynomial})
            with $n_{\rm max}=3$.
\end{enumerate}
  Finally, we calculated the imaginary
  part of up- and strange-quark number densities,
  $\textrm{Im}(n_{u})$ and $\textrm{Im}(n_{s})$. 
  We consider the situation that 
  two of chemical potentials are fixed to constant values
  and thereby the RW periodicity disappears.
  When $\theta_{s}=0$ and $\theta_{d}\le \pi/8$,
  $\textrm{Im}(n_{u})$ becomes an analytic function of $\theta_{u}$ 
  for any $T$. 
  The condition for $\textrm{Im}\left(n_{s}\right)$ to be 
  an analytic function of $\theta_{s}$ for any $T$ is 
  $\theta_{u}=\theta_{d}=\theta_{l}\le \pi/5$.
  When these conditions are satisfied, the values of $n_{u}$ and $n_{s}$ 
  can be calculated by the analytic continuation, for any $T$ of interest. 

  In the present paper, we concentrate on qualitative discussion based on
  the extended $\mathbb{Z}_{3}$ symmetry.
  The results mentioned above are interesting theoretically,
  but do not exactly correspond to the realistic case.
  As a future work, it is quite interesting to make systematic
  and quantitative analyzes, particularly in more realistic cases.

\noindent
\begin{acknowledgements}
Authors thank K. Kashiwa, M. Ishii and S. Togawa for useful discussions and comments.
J. S., H. K., and M. Y. are supported
by Grant-in-Aid for Scientific Research (No. 27-7804, No. 26400279, and No. 26400278)
from the Japan Society for the Promotion of Science (JSPS). 
\end{acknowledgements}


\end{document}